\documentclass[slac_one]{revtex4}
\usepackage{graphicx}
\usepackage{fancyhdr}
\usepackage{subfigure}
\def\W{$W$}
\def\Z{$Z$}
\def\Zgeq{Z/\gamma^*}
\def\Zg{$\Zgeq$}
\def\Zgint{$Z-\gamma^*$}

\def\gW{$\Gamma_{W}$}

\def\mTeq{M_{T}}
\def\mT{$\mTeq$}
\def\pt{$p_{T}$}
\def\ptz{$p_{T}^{Z}$}
\def\rapZeq{y_{Z}}
\def\rapZ{$\rapZeq$}
\def\ptleq{p_{T}^{\ell}}
\def\ptl{$\ptleq$}

\def\Wenu{$W\rightarrow e\nu$}
\def\Wmunu{$W\rightarrow\mu\nu$}
\def\Wlnu{$W\rightarrow\ell\nu$}

\def\Zll{$Z\rightarrow\ell\ell$}
\def\Zee{$Z\rightarrow ee$}

\def\AFB{$A_{FB}$}
\def\sint{$\rm sin^2\theta_W^{eff}$}
\begin{document}

\title{{\small{Hadron Collider Physics Symposium (HCP2008),
Galena, Illinois, USA}}\\ 
\vspace{12pt}
W and Z properties at the Tevatron} 

%

\author{Emily Nurse (for the CDF and D\O\ collaborations).}
\affiliation{University College London, London, UK.}

\begin{abstract}
I present measurements of \W\ and \Z\ boson properties by the CDF and D\O\ collaborations.
This includes measurements that test the production mechanism of the bosons and precision measurements of 
electroweak parameters. 
In the former category I present CDF measurements of the \Z\ rapidity and \W\ charge
asymmetry that will help further constrain Parton Distribution Functions in future global fits, and a D\O\
measurement of the \Z\ transverse momentum distribution that can be used to test the predictions of quantum-chromodynamics.
In the later category I present a D\O\ measurement of the \Zg\ forward-backward
asymmetry and the subsequent extraction of \mbox{\sint\ = 0.2327 $\pm$ 0.0018 (stat.) $\pm$ 0.0006 (syst.)} and a CDF measurement of the \W\ width
(\gW) using a fit to the tail  of the \W\ transverse mass distribution in \Wenu\ and \Wmunu\ events that yields \mbox{\gW\ = 2032 $\pm$ 73~MeV}.
\end{abstract}
\maketitle

\thispagestyle{fancy}

\section{Introduction}
\W\ and \Z\ bosons are produced at the Tevatron proton-antiproton collider via quark-antiquark annihilation. 
They quickly decay into two fermions which, due to the large mass of the bosons,
 typically have a high momentum in the direction transverse to the beam, \pt.
The \Zll\ and \Wlnu\ decay channels, where the charged lepton is an electron or muon, 
have clean experimental signatures and can therefore be utilized to make precision measurements of \W\ and \Z\ properties
and to probe the quantum-chromodynamic (QCD) aspects of their production mechanism.
\Z\ events are identified by the detection of two high \pt\ charged leptons ($\mu^+\mu^-$ or $e^+e^-$)
\footnote{It should be noted that the exchange of a virtual photon and \Zgint\ interference terms are indistinguishable from pure \Z\ exchange.
           For the remainder of this document I shall use \Z\ to represent \Z\ or $\gamma^*$ exchange or interference between them,
           where the invariant mass of the decay products is close to the \Z\ mass to maximize the contribution from pure \Z\ exchange.}. 
Both electrons and muons are detected in the CDF and D\O\ central trackers where their momenta are measured. 
Muons tracks are also matched to tracks in the outer muon detectors.
Electrons are also detected in the calorimeters, where a measurement of their energy is made.
\W\ events are identified by one high \pt\ charged lepton and one high \pt\ neutrino. Again, the charged lepton
is detected and its momentum measured, but the neutrino cannot be detected. Its existence is inferred by an imbalance
in \pt\ in the detectors.

The remainder of this document is organized as follows.
Section~\ref{sec-PDF} describes how \W\ and \Z\ events can place constraints on Parton Distribution Functions (PDFs) and 
presents CDF measurements of the \Z\ rapidity distribution and the \W\ charge asymmetry. Section~\ref{sec-zpt} motivates and 
describes a D\O\ measurement of the \Z\ \pt\ distribution. Section~\ref{sec-zasym} presents a D\O\ measurement of the \Zg\
forward-backward asymmetry and extraction of \sint\ and Section~\ref{sec-width} presents a precision measurement
of the \W\ width from CDF.
All of the presented measurements have been corrected for detector acceptance and smearing affects and can be compared directly
to theoretical predictions.
\section{\boldmath PDF constraints with \W\ and \Z\ events}
\label{sec-PDF}
PDFs are parameterized functional forms that describe the momentum distribution of partons in hadrons.
The parameters are constrained by fits to many data sets from different experiments.
The PDF sets used in these analyses are from global fits performed by the CTEQ and MRST groups.

Well constrained PDFs are essential for many measurements and searches at hadron colliders since they affect the simulated cross-sections
and kinematic distributions of signal and background predictions from Monte Carlo event generators and theoretical calculations.
At leading-order the \W\ and \Z\ production cross-sections can be written as:
\begin{equation}\label{eq:sigma}
\sigma_{p\bar{p}\rightarrow W/Z}
 =  \int \Sigma_{i,j} [f_i^q(x_p)f_j^{\bar{q}}(x_{\bar{p}}) + f_i^{\bar{q}}(x_p)f_j^{q}(x_{\bar{p}})]
\times d\sigma_{q\bar{q}\rightarrow W/Z}dx_{p}dx_{\bar{p}}
\end{equation}
where $f_i^q(x_p)$ gives the probability of a quark of flavor $i$ to be carrying a fraction $x_p$ of the proton's momentum (this is the PDF for parton $i$).
$f_j^{\bar{q}}(x_{\bar{p}})$, $f_i^{\bar{q}}(x_p)$ and $f_j^{q}(x_{\bar{p}})$ give the PDFs for antiquarks in the antiproton, antiquarks in the proton and
quarks in the antiproton respectively.
$d\sigma_{q\bar{q}\rightarrow W/Z}$ is the cross-section of the hard scatter for a given $x_p$ and $x_{\bar{p}}$.
Differential measurements of the cross-sections with respect to any variables that depend on $x_p$, $x_{\bar{p}}$ or the parton flavor can therefore be used to constrain PDFs. 
\subsection{\boldmath Measurement of the \Z\ rapidity distribution}
In \Z\ production $x_p$ and $x_{\bar{p}}$ are related to the rapidity of the \Z\ boson, \rapZ, via the equations:
\begin{equation}\label{eq:zrapA}
x_{p} = M_Ze^{\rapZeq}s^{-\frac{1}{2}}
\end{equation}
\begin{equation}\label{eq:zrapB}
x_{\bar{p}} = M_Ze^{-\rapZeq}s^{-\frac{1}{2}}
\end{equation}
where $s$ is the center of mass energy and $M_Z$ is the mass of the \Z.
Thus a measurement of $d\sigma(Z)/dy$ places constraints on the proton PDFs.
In particular, the high  $|\rapZeq|$ region probes both the high and low $x$ regions of the PDFs, as can be deduced from equations~\ref{eq:zrapA} and~\ref{eq:zrapB}.

\begin{figure*}[t]
\centering
\includegraphics[width=80mm]{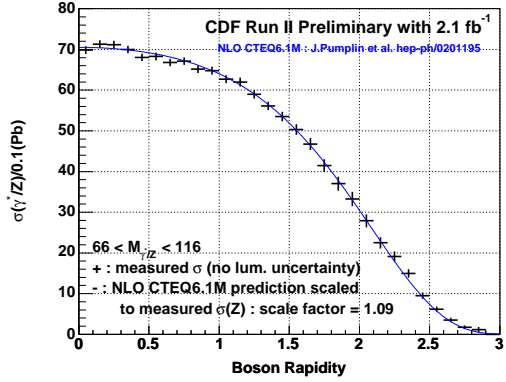}
\caption{The measurement of  $d\sigma(Z)/dy$ compared to a NLO calculation using NLO CTEQ6.1M PDFs.} \label{FIG:Zrap}
\end{figure*}

CDF has made a measurement of $d\sigma(Z)/dy$ using 2.1 fb$^{-1}$ of  \Zee\ data with $|\eta_e| < 2.8$.
Figure~\ref{FIG:Zrap} shows the measurement compared to a next-to-leading-order (NLO) calculation~\cite{ref:nlo} using NLO CTEQ6.1M~\cite{ref:cteq-nlo} PDFs.
Good agreement is observed after scaling the normalization of the prediction to match the data.
The uncertainties on the data points include both statistical and systematic uncertainties, but exclude an overall luminosity uncertainty. 
\begin{figure*}[t]
\centering
\subfigure[]{
\includegraphics[width=85mm]{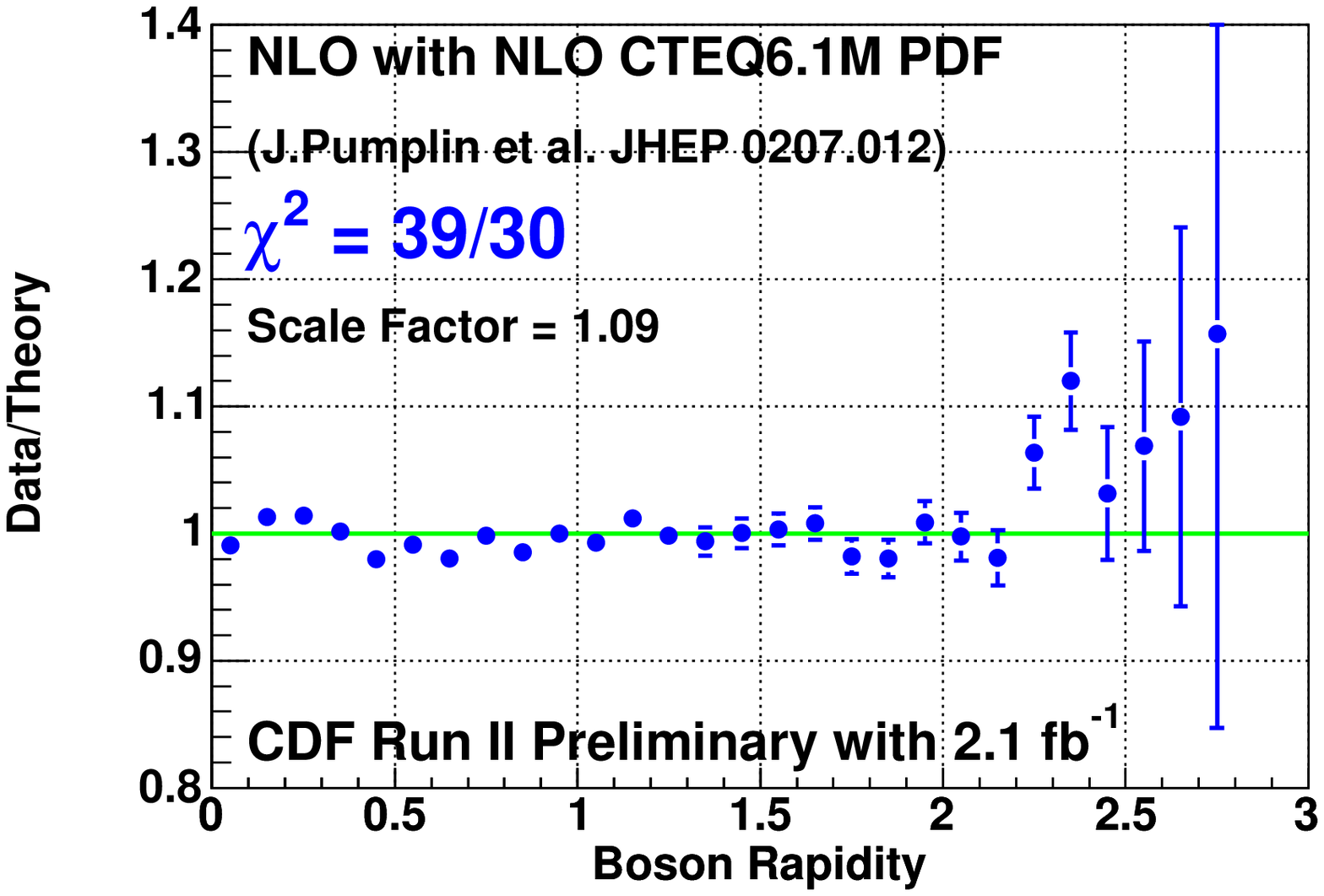}
}
\subfigure[]{
\includegraphics[width=85mm]{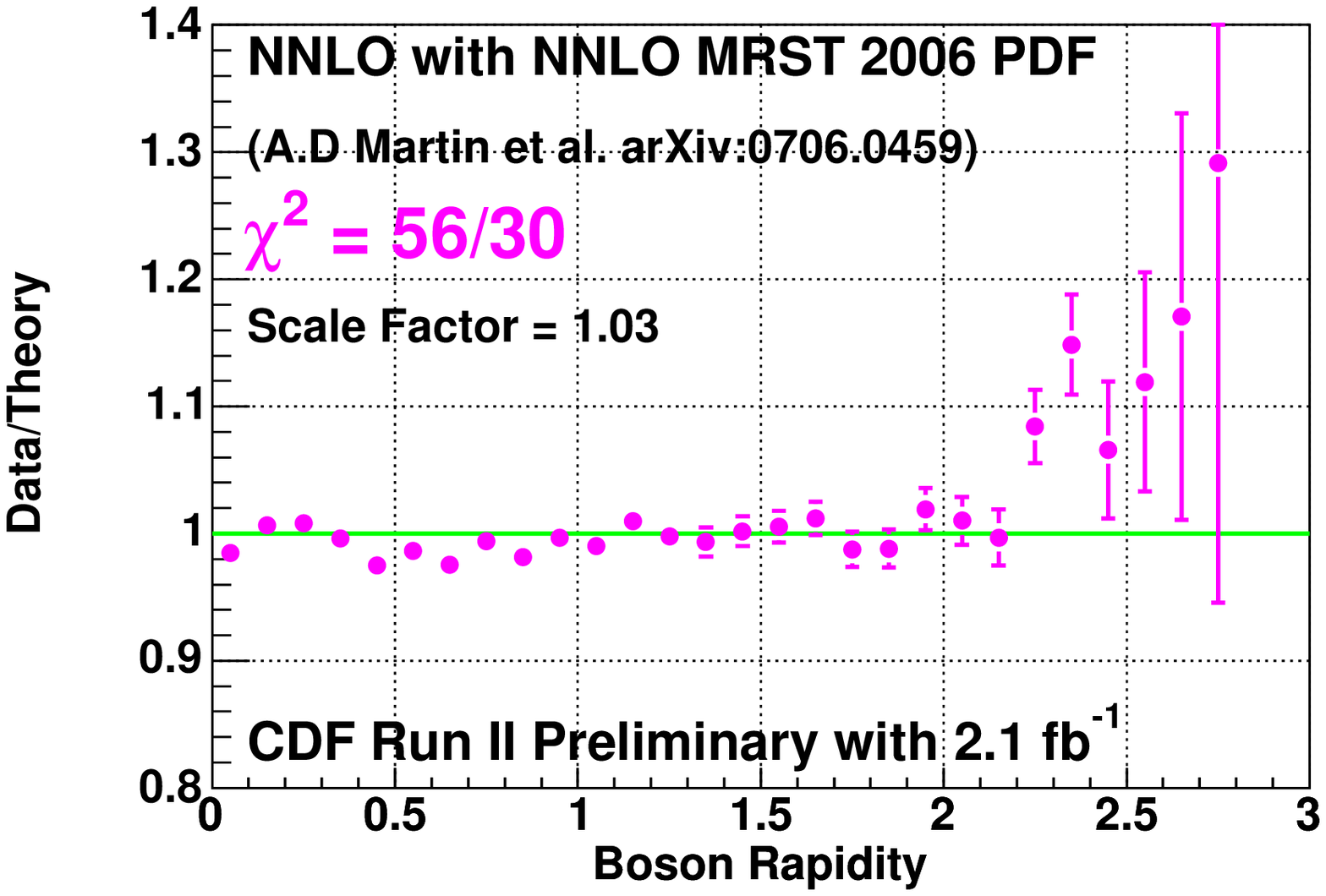}
}
\caption{Data divided by theory for $d\sigma(Z)/dy$ for (a) a NLO calculation using NLO CTEQ6.1M PDFs and (b) a NNLO calculation using NNLO MRST 2006 PDFs.} 
\label{FIG:Zrapratio}
\end{figure*}
Figure~\ref{FIG:Zrapratio} shows the distribution of the data divided by the theory for (a) the NLO calculation with NLO CTEQ6.1M PDFs and 
(b) a next-to-next-to-leading-order (NNLO) calculation~\cite{ref:nnlo} using NNLO MRST 2006 PDFs~\cite{ref:mrst-nnlo}.
Only the statistical errors are included in these ratio plots as the systematics have large bin to bin correlations.
The data are found to be most consistent with the NLO CTEQ6.1M prediction. This result will help to constrain PDFs in future fits.
\subsection{\boldmath Measurement of the \W\ charge asymmetry}
On average the $u(\bar{u})$ quark carries a higher
fraction of the (anti)proton's momentum than the $d(\bar{d})$ quark, meaning that a $W^+(W^-)$ produced  via $u\bar{d}(d\bar{u})$ 
annihilation will tend to be boosted in the direction of the (anti)proton beam.
This results in a \W\ charge asymmetry, defined as:
\begin{equation}
A(y_{W}) = \frac{d\sigma(W^{+})/dy_{W} - d\sigma(W^{-})/dy_{W}}{d\sigma(W^{+})/dy_{W} + d\sigma(W^{-})/dy_{W}},
\end{equation}
where $y_{W}$ is the rapidity of the \W\ and $\sigma(W^{\pm})$ are the cross-sections for $W^{\pm}$ production.
This distribution is sensitive to the ratio of the $u$ and $d$ PDFs at the scale $Q^2 \approx M_{W}^{2}$
and is an important input to the global PDF fits.

Since the neutrino momentum in the direction of the beam, $p_{z}^{\nu}$, is not known, a measurement of the 
electron or muon charge asymmetry, as opposed to $A(y_{W})$, has traditionally been made. 
This distribution is a convolution of $A(y_{W})$ and the \Wlnu\ angular decay structure.
The fact that the \W\ only couples to left(right) handed (anti)particles favors decay to a forward(backward) $\ell^{-(+)}$. 
This dilutes the effect of the production asymmetry and reduces the constraint on the $u:d$ ratio  provided by the measurement.

\begin{figure*}[t]
\centering
\subfigure[]{
\includegraphics[width=80mm]{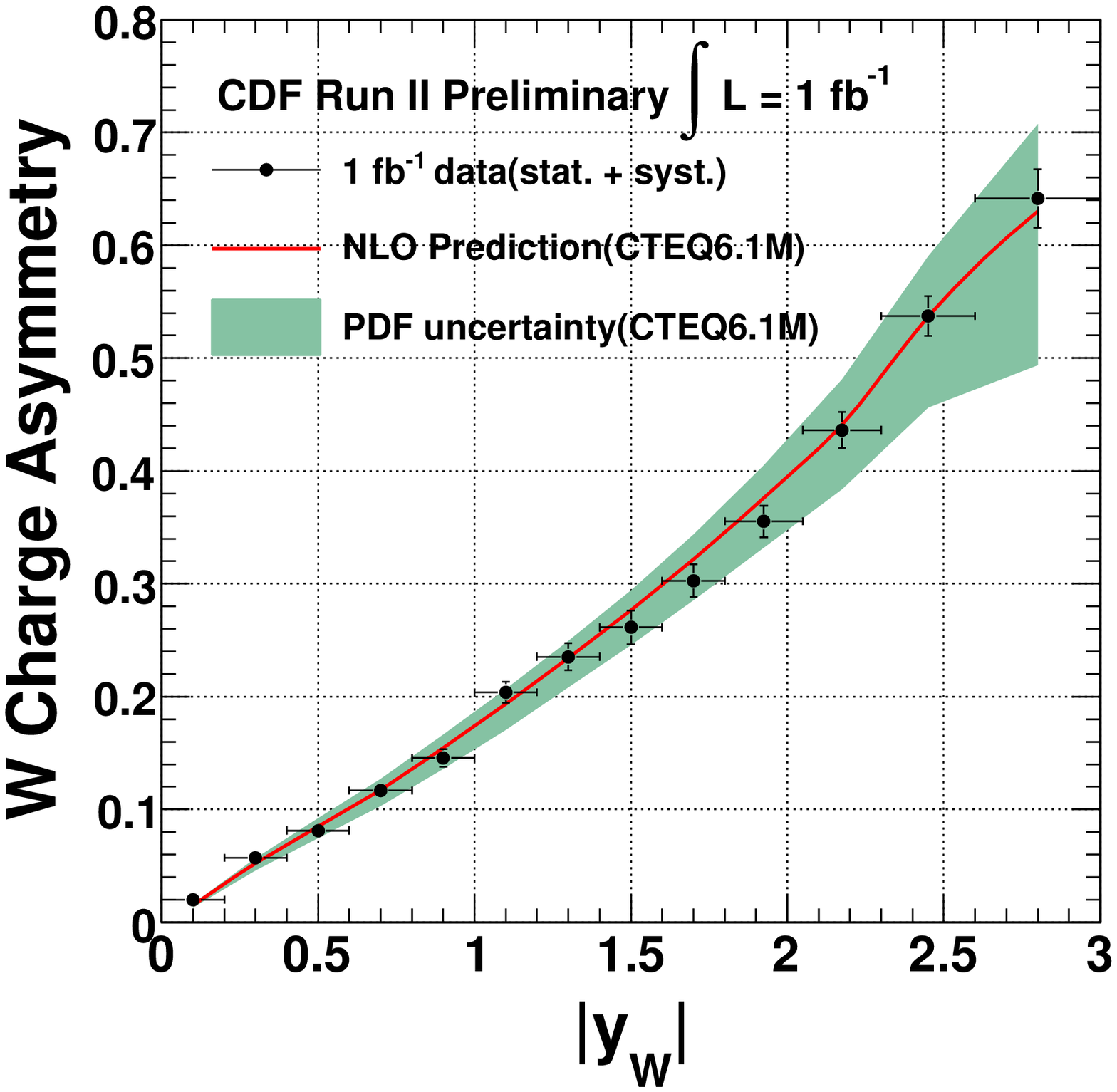}
}
\subfigure[]{
\includegraphics[width=80mm]{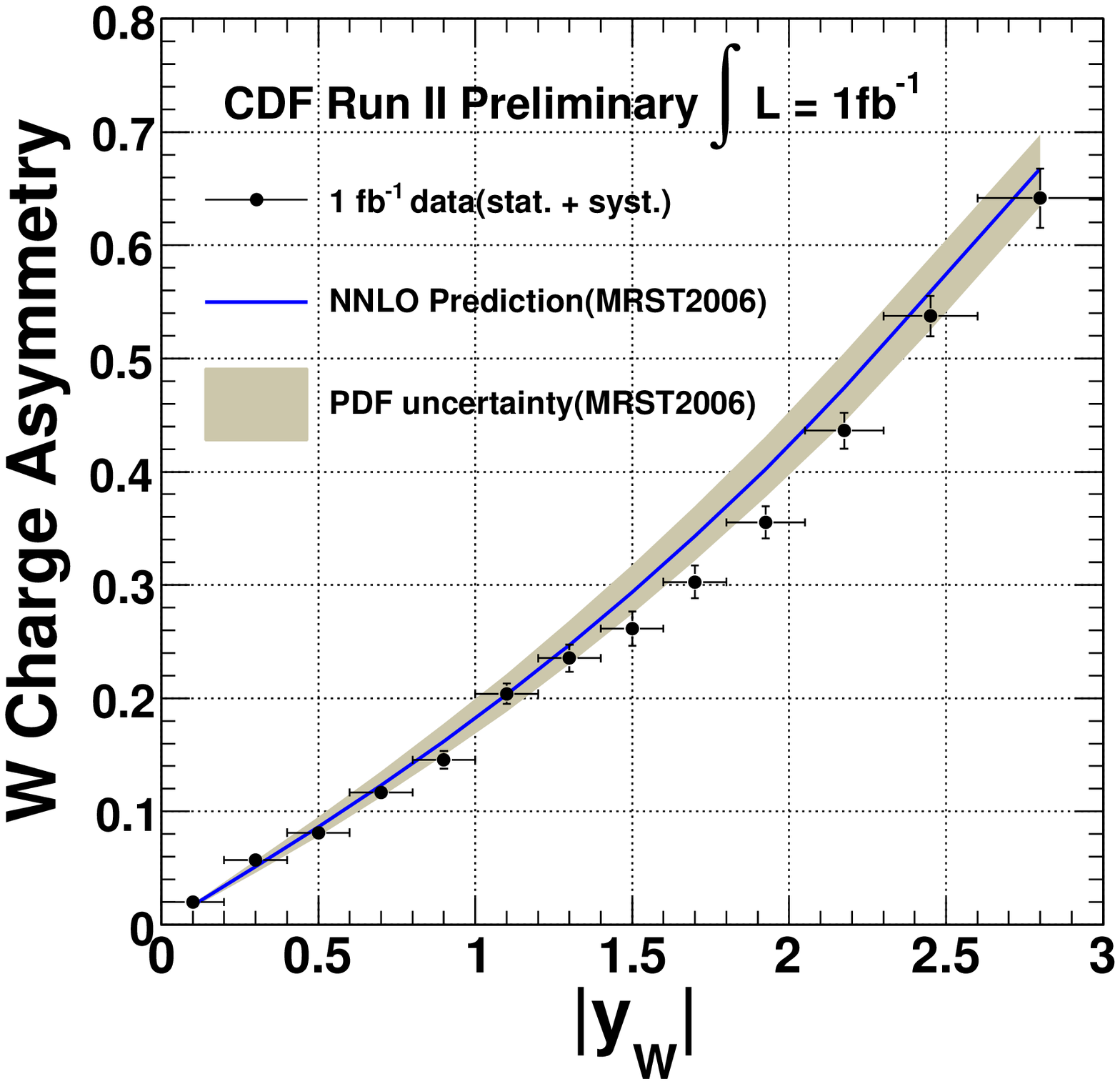}
}
\caption{The \W\ charge asymmetry measurement compared to (a) NLO CTEQ6.1M and (b)  NNLO MRST 2006 PDFs and their uncertainties.} \label{FIG:Wasym}
\end{figure*}

The CDF analysis extracts $A(y_{W})$ by constraining $M_{W} = 80.4~\rm GeV/c^2$, giving two possible solutions
for $p_{z}^{\nu}$. Each solution receives a probability weight according to the  decay structure 
and $\sigma(W^{\pm})$. The \W\ charge asymmetry is then extracted after correcting for detector effects.
The process is iterated to reduce the dependence of the weighting factor on the asymmetry itself.
The measured \W\ charge asymmetry is compared to the prediction from PDFs using (a) NLO CTEQ6.1M~\cite{ref:cteq-nlo} and
(b) NNLO MRST 2006~\cite{ref:mrst-nnlo}  in Figure~\ref{FIG:Wasym}. 
The data are most consistent with the NLO CTEQ6.1M prediction.
The PDF uncertainties from the respective groups are shown with shaded bands.
The experimental uncertainties are smaller than the uncertainties from the PDFs indicating that this measurement will help
to constrain PDFs in future fits.
\section{\boldmath Measurement of the \Z\ \pt\ distribution}
\label{sec-zpt}
At leading-order \Z\ bosons are produced with zero \pt\ because the annihilating quarks are moving in the 
direction of the incoming (anti)protons. Higher order QCD corrections lead to parton radiation from the incoming quarks
which give the \Z\ a non-zero \pt. 
Therefore a measurement of the \Z\ \pt\ (\ptz) distribution provides a stringent test of QCD predictions and can be used 
to tune and validate Monte Carlo event generators.
At large \ptz\ (greater than about \mbox{30 GeV/c}) the radiation of a single (or double) parton dominates the cross-section and fixed 
order perturbative QCD calculations should yield reliable results. At lower \ptz\ multiple soft parton radiation dominates 
and resummation techniques or parton shower Monte Carlo event generators combined with non-perturbative models are required
to give reliable predictions.

\begin{figure*}[t]
\centering
\subfigure[]{
\includegraphics[width=80mm]{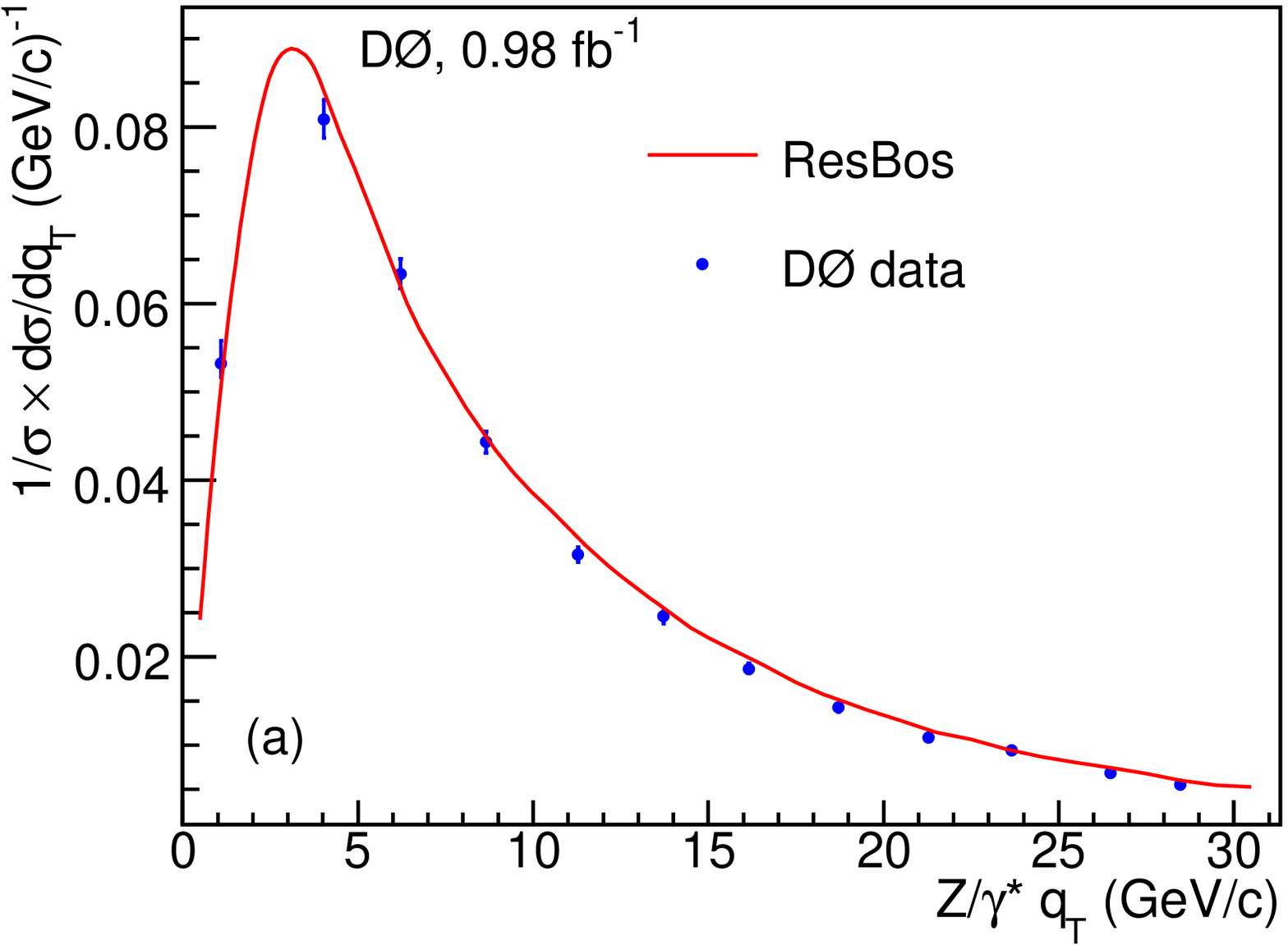}
\label{FIG:ZpT}
}
\subfigure[]{
\includegraphics[width=80mm]{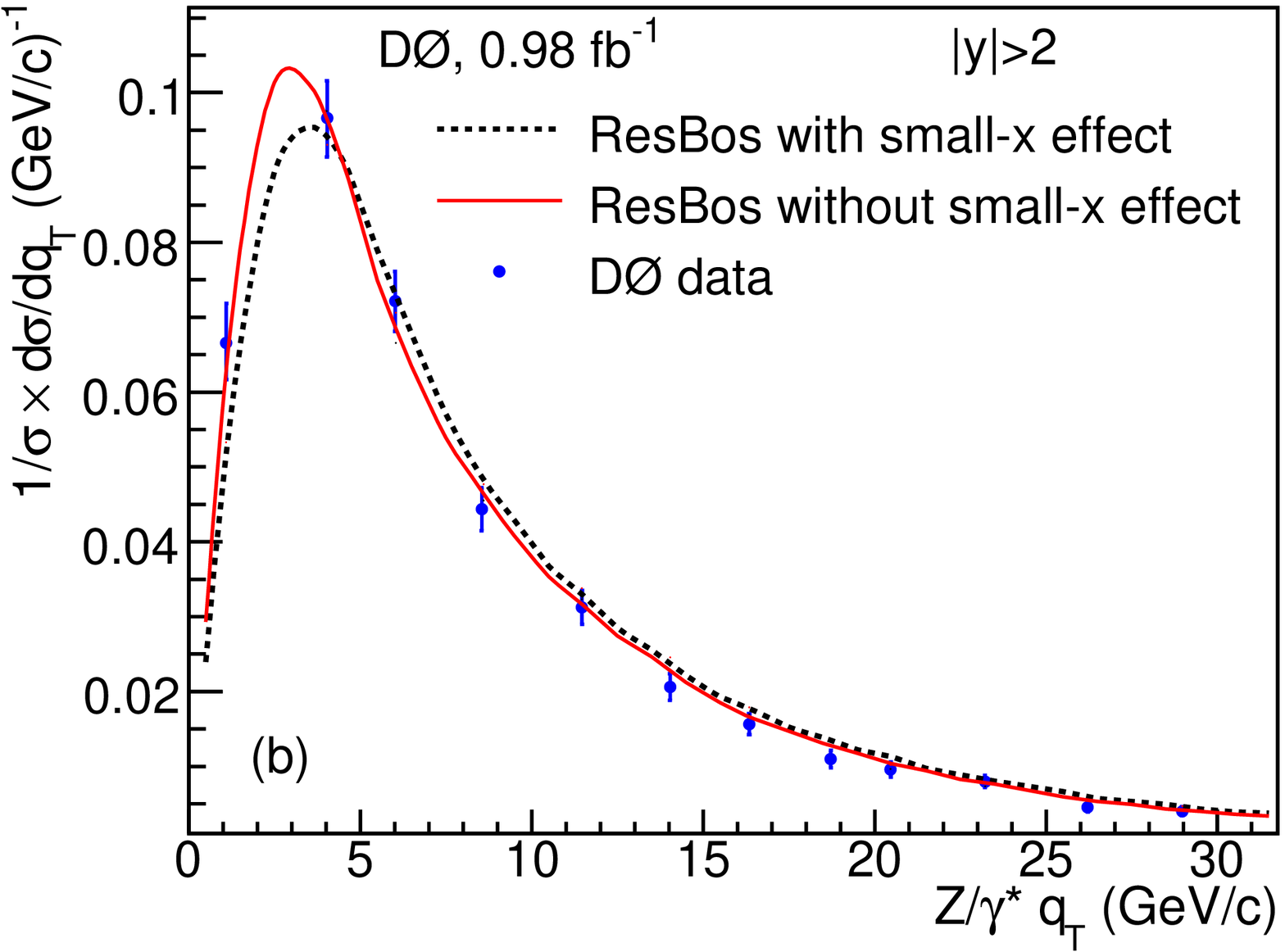}
\label{FIG:ZpTlowx}
}
\caption{The normalized differential cross-section as a function of \pt\ (denoted $q_{T}$ in the plot) in the region \pt\ $<$ 30 GeV/c for (a) the inclusive sample and 
         (b) the sample with $|$\rapZ$| >$ 2. The data are compared to the prediction from ResBos.}
\end{figure*}

D\O\ have published a measurement of the normalized differential cross-section as a function of \ptz\  using $0.98 \rm ~fb^{-1}$ of \Zee\ data with $|\eta_{e}| < 3.2$~\cite{ref:D0-zpt}.
Figure~\ref{FIG:ZpT} shows the result to be in good agreement with the prediction from the ResBos~\cite{ref:resbos} event generator in the region \ptz\ $<$ 30~GeV/c.  
ResBos incorporates a NLO perturbative QCD calculation at high \ptz\ with the Collins, Soper and Sterman (CSS)~\cite{ref:css} resummation formalism in impact parameter space
using the Brock, Landry, Nadolsky and Yuan (BNLY)~\cite{ref:bnly} non-perturbative function.

Recent studies from deep inelastic scattering experiments~\cite{ref:dis} indicate that the resummation calculation may need to be modified for
\Z\ bosons produced with a small-$x$ parton. The \ptz\ distribution is expected to broaden in the small-$x$ (high $|$\rapZ$|$) region which has important implications 
for modeling vector boson and Higgs production at the LHC~\cite{ref:smallx}.
Figure~\ref{FIG:ZpTlowx} shows the \ptz\ distribution for events with $|$\rapZ$| >$ 2 compared to ResBos with and without the modified calculation.
A better description is currently obtained from ResBos without the modified calculation, although it remains to be seen whether a re-tuning of the
non-perturbative parameters could improve the agreement.

\begin{figure*}[t]
\centering
\subfigure[]{
\includegraphics[width=80mm]{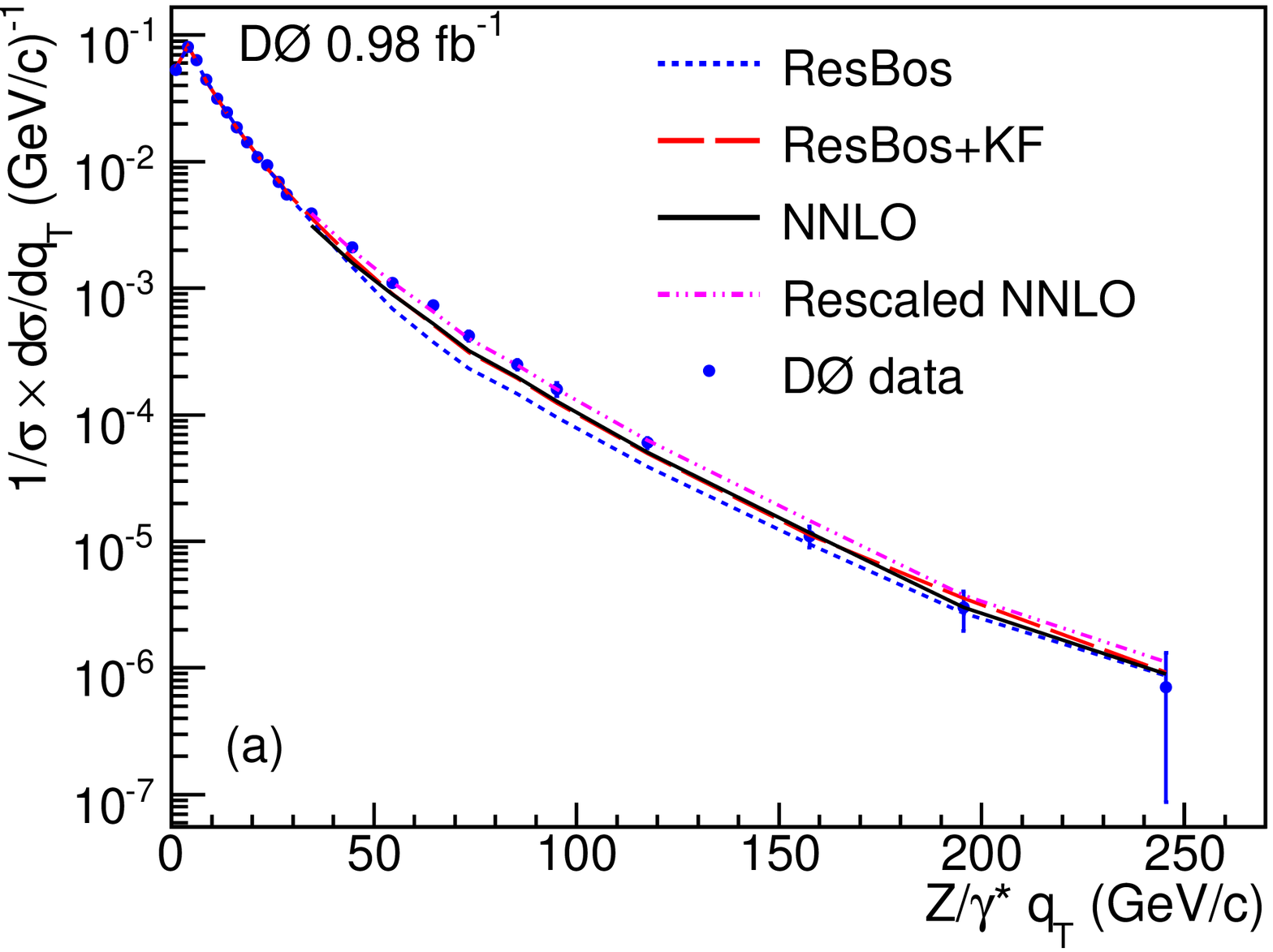}
 \label{FIG:ZpTwide}
}
\subfigure[]{
\includegraphics[width=80mm]{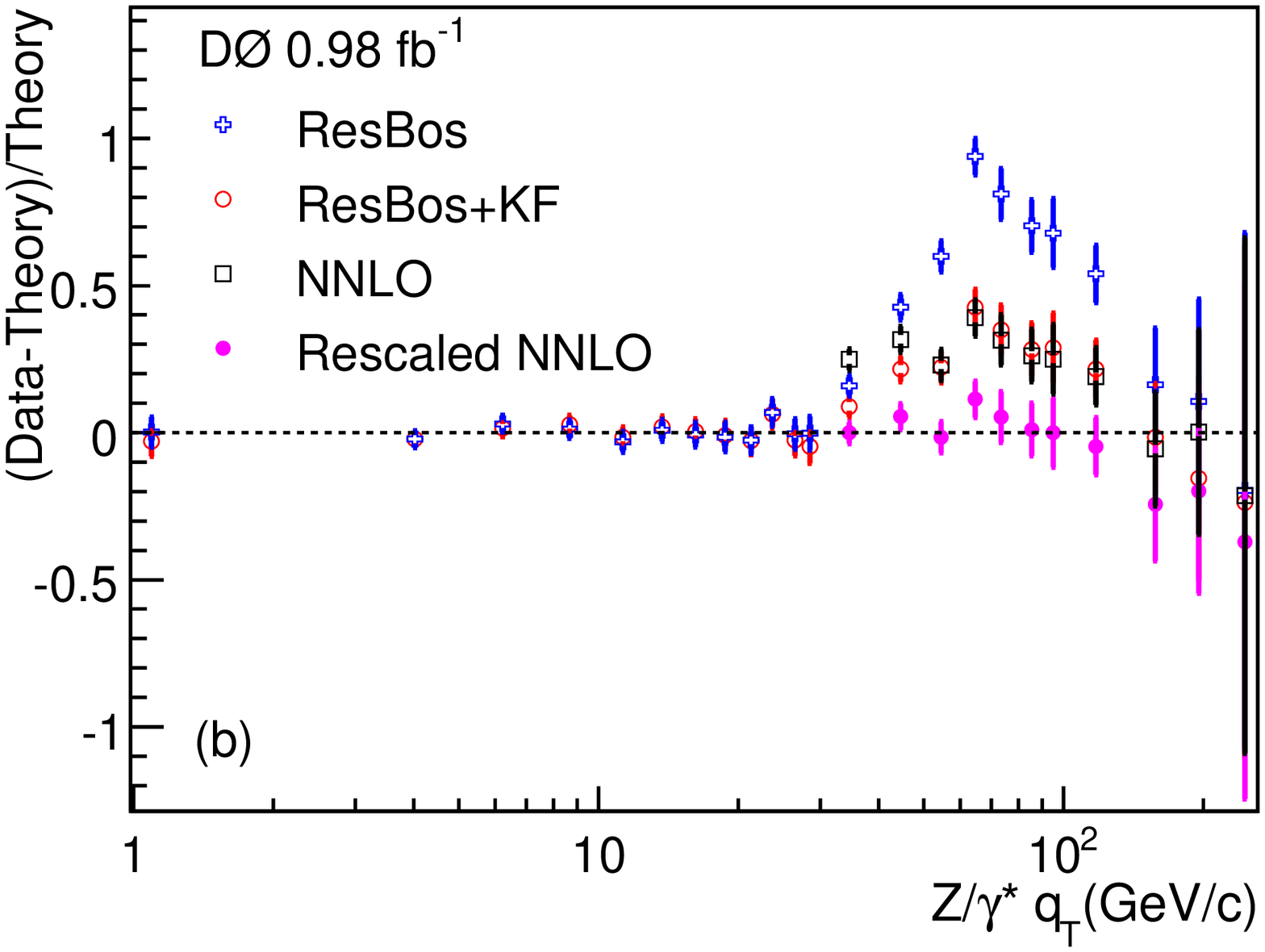}
\label{FIG:ZpTwide-ratio}
}
\caption{(a) The normalized differential cross-section as a function of \pt\ (denoted $q_{T}$ in the plot) compared to four theoretical predictions:
         (1) ResBos, (2) ResBos with a NLO to NNLO K-factor, (3) a NNLO calculation and (4) the NNLO calculation but rescaled to the 
         data at \pt\ = 30 GeV/c.
         (b) Data minus theory, divided by theory for the above predictions.} 
\end{figure*}

Figure~\ref{FIG:ZpTwide} shows the measured \ptz\ distribution in the region  \ptz\ $<$ 260~GeV/c compared to ResBos, 
ResBos with a NLO to NNLO K-factor provided by~\cite{ref:kfactor}, a NNLO perturbative calculation~\cite{ref:ZpT-NNLO} divided by a NNLO calculation 
of the total cross-section~\cite{ref:sigma-NNLO} and the NNLO calculation rescaled to match the data at  \ptz\ = 30~GeV/c.
Figure~\ref{FIG:ZpTwide-ratio} shows the data minus theory, divided by theory distributions for the above predictions.
In the large \ptz\ region ResBos under-estimates the cross-section. Applying the NLO to NNLO K-factor improves the prediction but the best agreement in the 
region \mbox{\pt\ $>$ 30~GeV/c} is achieved when the prediction is rescaled by a factor of 1.25 to match the data at \mbox{\ptz\ = 30~GeV/c.} This indicates that the shape of the
distribution is well modeled but the source of the discrepancy is in the normalization.
\section{\boldmath \Z\ forward-backward asymmetry}
\label{sec-zasym}
The \Z\ couplings to fermions have both vector and axial-vector components. 
The angular decay structure from vector couplings has the form $\frac{d\sigma}{d \rm cos \theta^*} \sim 1 + \rm cos^2\theta^*$ and that from 
axial-vector couplings has the form $\frac{d\sigma}{d \rm cos \theta^*} \sim \rm cos\theta^*$, where $\theta^*$ is the angle between the outgoing 
negatively charged lepton and the incoming quark in the rest frame of the lepton pair. 
The axial-vector contribution gives rise to a forward-backward asymmetry defined as:
\begin{equation}
A_{FB} = \frac{\sigma_F - \sigma_B}{\sigma_F + \sigma_B}
\end{equation}
where $\sigma_F$ is the forward cross-section (for events with  cos$\theta^* > 0$) and 
$\sigma_B$ is the backward cross-section (for events with cos$\theta^* < 0$).
Since \AFB\ depends on the relative size of the vector and axial-vector couplings it is sensitive to
\sint\ where $\rm \theta_W^{eff}$ is the effective weak mixing angle.
\AFB\ varies with the di-lepton invariant mass, $M_{\ell\ell}$, as the relative contributions from the exchange of \Z, $\gamma^*$ and the interference terms vary. 
A measurement of \AFB\ as a function of $M_{\ell\ell}$ can be used to extract \sint.

\begin{figure*}[t]
\centering
\includegraphics[width=100mm]{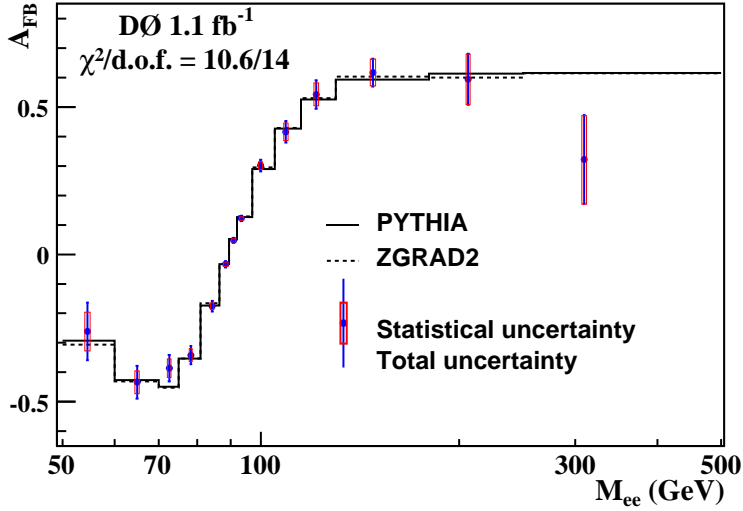}
\caption{The unfolded \AFB\ as a function of the di-electron invariant mass compared to the prediction from Pythia and ZGRAD2.}
\label{FIG:AFB}
\end{figure*}

D\O\ has published a measurement of \AFB\ using 1.1 fb$^{-1}$ of \Zee\ data with $|\eta_e| < 2.5$~\cite{ref:D0-afb}. 
Figure~\ref{FIG:AFB} shows the unfolded \AFB\ as a function of the di-electron invariant mass compared to Standard Model predictions from
the Pythia~\cite{ref:pythia} and ZGRAD2~\cite{ref:zgrad2} event generators.
The data are in good agreement with the Standard Model prediction. It should be noted that significant deviations from the Standard Model
in the high mass region could signify the existence of a new massive neutral gauge boson.
A fit to the \AFB\ distribution yields a measurement of \mbox{\sint\ = 0.2327 $\pm$ 0.0018 (stat.) $\pm$ 0.0006 (syst.)}, 
which is in good agreement with previous measurements~\cite{ref:pdg}.
\section{\boldmath \W\ width measurement}
\label{sec-width}
The \W\ width is known within the Standard Model to an extremely high precision of 0.1\%, thus an accurate experimental measurement
is desirable to test this prediction.

CDF has published a measurement of \gW\ using 350~pb$^{-1}$ of \Wmunu\ and \Wenu\ data~\cite{ref:CDF-width}.
Since neutrinos are not detected in CDF, the invariant mass of the \W\ decay products cannot be reconstructed. 
Instead we measure \gW\ by reconstructing the transverse mass, which is defined as  $\mTeq = \sqrt{2\ptleq p_{T}^{\nu}(1 - \cos \phi_{\ell\nu})}$,
where \ptl\ and  $p_{T}^{\nu}$ are the \pt\ of the charged lepton and the neutrino respectively
and $\phi_{\ell\nu}$ is the azimuthal angle between them.
$p_{T}^{\nu}$ is inferred from the transverse momentum imbalance in the event.

A fast, parameterized Monte Carlo simulation is used to predict the \mT\ distribution as a function of \gW.
After adding background distributions to the Monte Carlo, a binned maximum-likelihood fit to the data is performed in the region 90~$<$~\mT~$<$~200~GeV/$c^2$ to extract \gW.
The fit is performed in the high \mT\ tail region because it is still sensitive to the Breit-Wigner line-shape but less
sensitive to the Gaussian detector resolutions than the peak region.
These line-shape predictions depend on a number of production and detector effects that must be accurately modeled.

\begin{figure*}[t]
\centering
\subfigure[]{
\includegraphics[width=80mm]{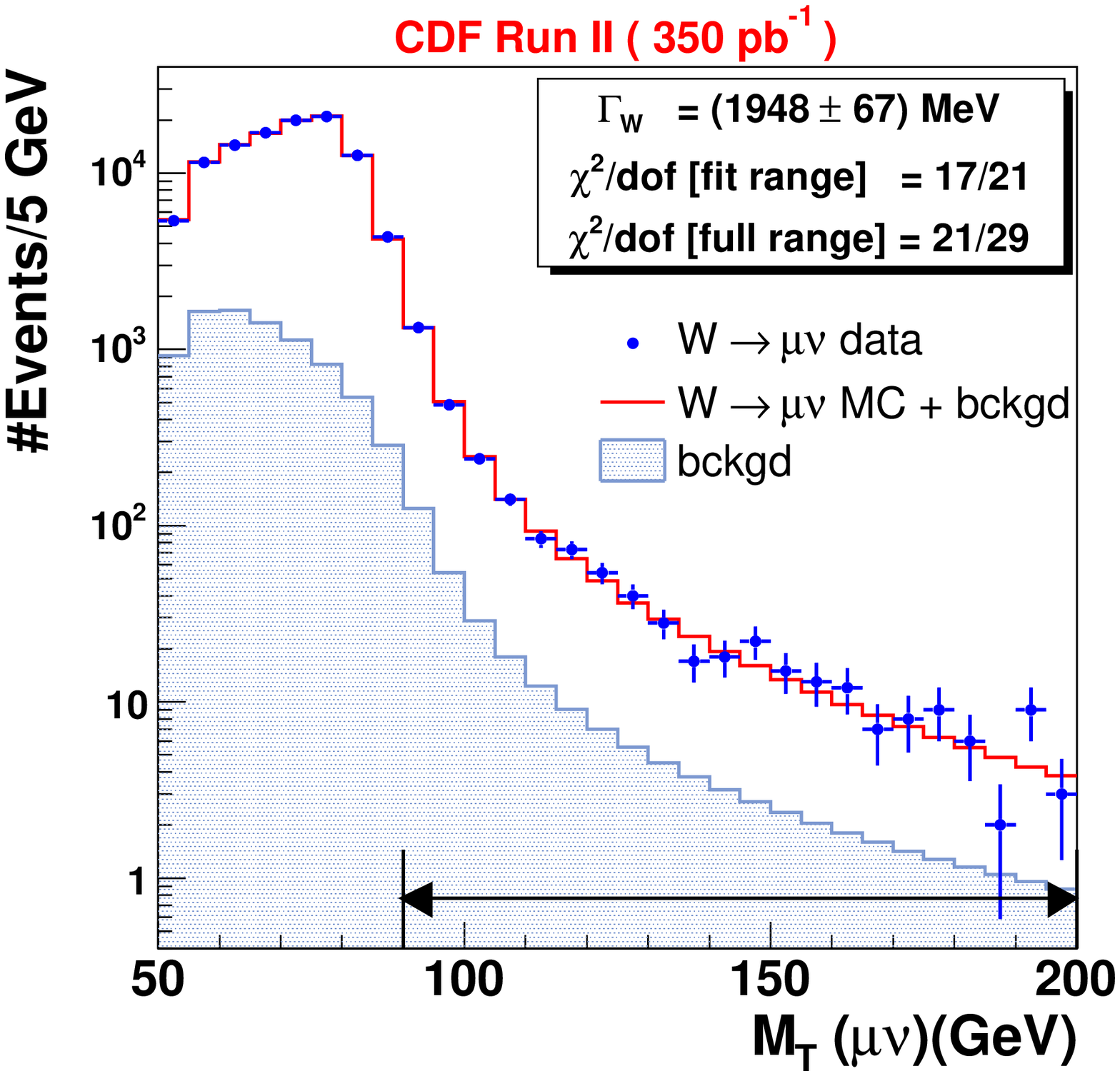}
}
\subfigure[]{
\includegraphics[width=80mm]{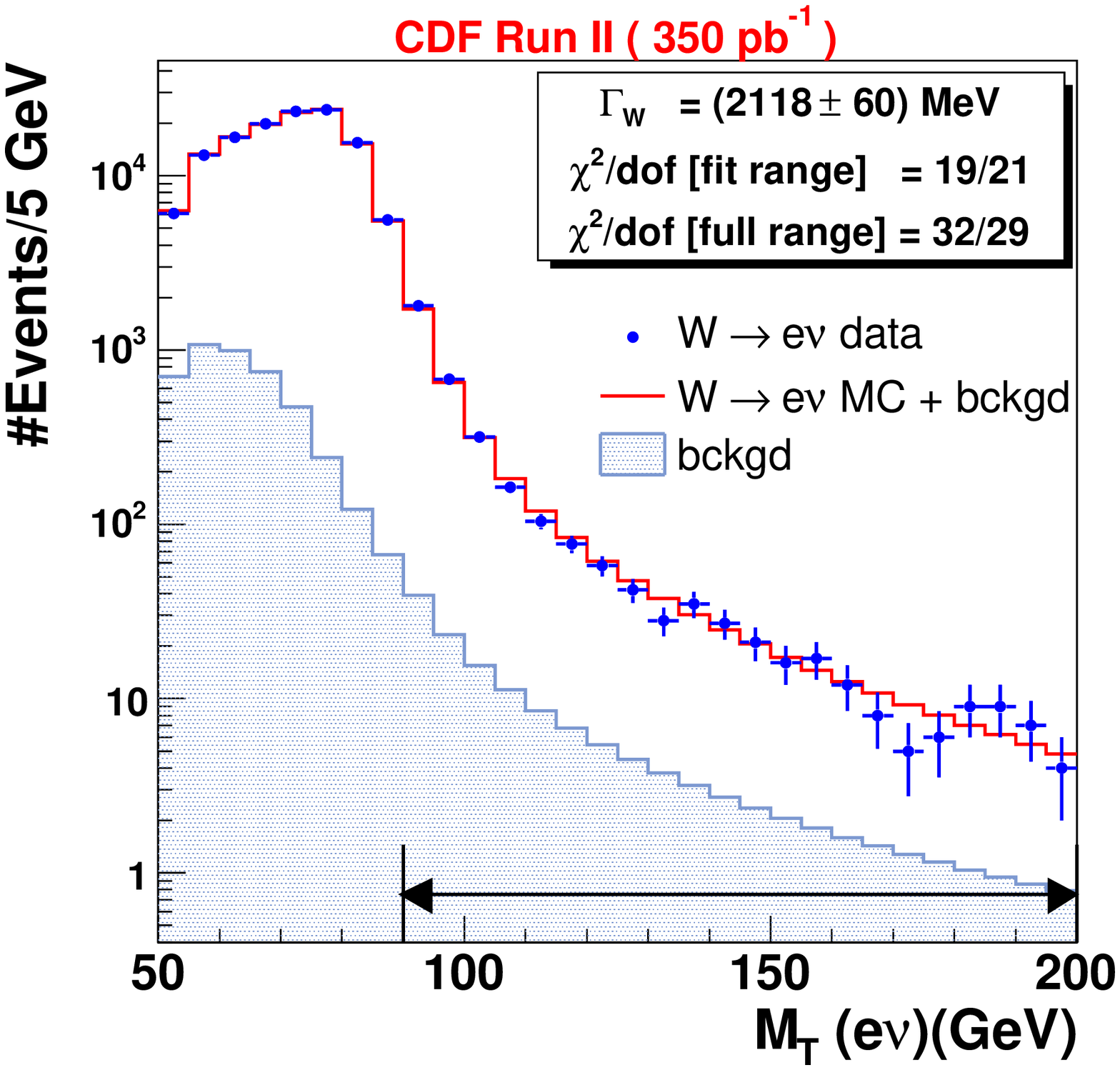}
}
\caption{The \mT\ distributions for (a) \Wmunu\ and (b) \Wenu\ data compared to the best fit Monte Carlo.} \label{FIG:width-fits}
\end{figure*}

Figure~\ref{FIG:width-fits} shows the  \mT\ fits for \gW\ in (a) \Wmunu\ and (b) \Wenu\ events. The results are combined to give the final result 
\gW\ = 2032 $\pm$ 73~MeV, the world's most precise  measurement, which is in good agreement with the Standard Model prediction of 2090.2 $\pm$ 0.9~MeV~\cite{ref:pdg}. This result reduces the
world average central value by 44~MeV and its uncertainty by 22\%. CDF has also made an indirect measurement of \gW, 
obtained from a measurement of the ratio of the cross-section times branching ratio for \Wlnu\ and \Zll~\cite{ref:indirect}.
This measurement yields \gW~=~2092~$\pm$~42~MeV, which is consistent with this direct measurement.
\section{Summary}
I have presented new measurements of the \Z\ rapidity distribution and \W\ charge asymmetry from CDF, which will be used in future global PDF fits. 
I have shown a new D\O\ measurement of the \Z\ \pt\ distribution that can be used to test QCD predictions and tune Monte Carlo event generators.
I have also presented a D\O\ measurement of the \Zg\ forward-backward asymmetry and subsequent extraction of \sint.
And lastly I have shown a direct measurement of the W width from CDF, which is the most precise in the world and in good agreement with the Standard Model.

\end{document}